\documentclass{article}
\usepackage[utf8]{inputenc}
\usepackage{amsfonts} 
\usepackage{amsmath} 
\usepackage{float}
\usepackage{graphicx}

\title{A model based on the fractional Brownian motion for the temperature fluctuation in the Campi Flegrei caldera}
\author{\small \textbf{Antonio Di Crescenzo$^{(1)}$, Barbara Martinucci$^{(1)}$ and Verdiana Mustaro$^{(1)}$} \\
\\
\footnotesize (1) Dipartimento di Matematica, Università di Salerno \\
\footnotesize I-84084 Fisciano (SA), Italy \; Email:\{adicrescenzo, bmartinucci, vmustaro\}@unisa.it}
\date{}

\begin{document}

\maketitle
\abstract{The aim of this research is to identify an efficient model to describe the fluctuations around the trend of the soil temperatures monitored in the volcanic caldera of the Campi Flegrei area in Naples (Italy). The study focuses on the data concerning the temperatures in the mentioned area through a seven-year period. The research is initially finalized to identify the deterministic component of the model, given by the seasonal trend of the temperatures, which is obtained through an adapted regression method on the time series. Subsequently, the stochastic component from the time series is tested to represent a fractional Brownian motion (fBm). An estimation based on the periodogram of the data is used to estabilish that the data series follows a fBm motion, rather then a fractional Gaussian noise. An estimation of the Hurst exponent $H$ of the process is also obtained. 
Finally, an inference test based on the detrended moving average of the data is adopted in order to assess the hypothesis that the time series follows a suitably estimated fBm.
}

\textit{Keywords: Fractional Brownian motion, stochastic model, regression, time series, residuals analysis, Hurst exponent, Campi Flegrei caldera, temperature fluctuation} 

\section{Introduction}
In several applied contexts, one of the main problems in the description of a phenomenon that evolves over time is the individuation of the probabilistic laws by which the phenomenon itself is driven. To this aim, schemes based on the superposition of deterministic and random components are often chosen, in which the deterministic part describes the phenomenon's trend, while the random one describes the fluctuations determined by unpredictable exogenous factors. An example of such model is presented in the work by Sebastiani and Malagnini \cite{sebastiani}, where a physical model with non-constant variance is proposed to describe the phenomenon of coupling erosion that caused the 2011 earthquake in Tohoku-Oki (Japan).
\par
The aim of the present study is to implement an analogous scheme to describe the soil temperatures observed on the surface of the Campi Flegrei caldera. This volcanic area is located near the city of Naples, in Italy, and is famous for the ground deformations that have been registered in the area since ancient times. Recent observations, carried out in the years from 2011 to 2017, were led to monitor the presence of radon in two sites of the area (cf. Sabbarese et al. \cite{0}). In the work \cite{0}, the presence of radon in the soil was tracked, along with its dependence on physical quantities such as temperature, pressure and humidity, in order to prove how the seismic activity in the Campi Flegrei area was influenced by the presence of the element. \par
Starting from the research mentioned above, in this work we aim to build a model able to describe the fluctuations of soil temperatures in the caldera through the identification of the temporal trend and the corresponding random component. \par 
What emerged from the present study is that the deterministic component is well described by a piecewise linear model, where the segments' slopes are alternating in sign. The slopes can be determined by applying the ordinary least squares (OLS) method on the observed data. The endpoints for each segment of the piecewise curve are chosen through an emphirical method. The stochastic model describing the random fluctuations component is instead recognized to be the fractional Brownian motion (fBm). For a description of the most relevant probabilistic and statistic aspects of fBm see, for example, Nourdin \cite{nourdin} and Prakasa-Rao \cite{rao}. This stochastic process has been proposed in several studies for modeling various geophysical phenomena. In the work by Mattia et al. \cite{mattia}, the fBm is proposed as an approximation of the process describing the trend of ground data involving soil roughness collected over three different European sites. In the paper by Yin and Ranalli \cite{yin}, the authors show that the event of earthquake rupturing in a fault is related to factors such as static shear stress and static frictional strength through the potential dynamic stress drop. The latter is regarded as a one-dimentional stochastic process and it is modeled as a fBm with Hurst exponent close to zero. \par
The evidence of reasonableness of the model firstly emerges as consequence of a statistical test, in which the hypothesis that the random fluctuations of the model can be described by a Brownian motion is rejected. The subsequent phase of the work involves the study of the periodogram of the data. It allows to verify that the random fluctuations are well described by a fBm, in opposition to a fractional Gaussian noise (fGn). Such analysis also leads to the estimation of the involved parameters, namely the Hurst exponent $H$ and the scale parameter $D$ of the fBm. Many estimation techniques for the former parameters have been proposed in the literature, such as the methods presented in \cite{taqqu2}. Other works often provide modified versions of the already known algorithms.\par
Subsequently, the study moves towards analysing the residual series emerging from the considered stochastic process and a simulated fBm with the previously estimated parameters. First of all, such residual series is plotted with respect to the temperatures series, which suggests the Gaussianity of the residuals. Then, the procedure is repeated through suitable simulations. Two statistical tests are performed on the obtained values, both leading to the acceptance of the null hypothesis of Gaussianity. \par
The goodness of the considered model is finally confirmed through an inference test in which the test statistic involves the detrending moving averages obtained from the data. \par
Here is the plan of the article. In Section 2 
we introduce the starting equation modeling the soil temperatures trend (cf. (\ref{model})). Section 3 focuses on the data analysis directed towards identifying the deterministic component of the temperatures trend. The remaining part of the article is dedicated to the analysis of the stochastic component of the model. More specifically, Section 4 is aimed to the study of the periodogram obtained from the data observations, from which emerges the eligibility of fBm to describe the random component of the temperature compared to fGn. An estimation of the relevant parameters of the process, i.e. the Hurst exponent and the scale parameter, is also performed. The testing on the residuals of the model is then performed by resorting to the Shapiro-Wilk test and the Robust Jarque-Bera test. The work is completed by Section 5, which contains the statistical test leading to the acceptance of the model, and Section 6, which collects the conclusions of the study.

\section{The stochastic model}
\label{sec:1}
The time series considered throughout this study describes the temperatures of the surface soil of the Campi Flegrei caldera, obtained via daily observations in the time period between 07/01/2011 and 12/31/2017. The main aim of the research is to develop a model able to describe the fluctuations of the data. As usually observed in geophysical studies, the trajectory of the time series is seen as the superposition of a deterministic component and a stochastic one, where the latter represents the fluctuations from the trend. As a consequence, the process $\{X(t), \; t \geq 0\}$ that represents the daily observed soil temperatures is described by the equation
\begin{equation} \label{model}
X(t)=r(t)+B_H(t),
\end{equation}
where $r(t)$ constitutes the deterministic component of the process, while $B_H(t)$ is a stochastic process. We will show that a suitable assumption is to identify $B_H(t)$ a fBm. A similar model has been proposed for the vertical motion of the soil in the Campi Flegrei area, cf. Travaglino et al. \cite{travaglino}. \par
Under the assumption that $B_H(t)$ is indeed a fBm, it is easy to observe that the model shown in (\ref{model}) is such that $\mathbb{E}(X(t))=r(t)$. We recall, in fact, that the fractional Brownian motion is a continuous-time Gaussian process with zero mean and covariance
\begin{equation} \label{eq2}
\textnormal{Cov}[B_H(t),B_H(s)]=D(t^{2H}+s^{2H}- |t-s|^{2H}), \quad t,s\geq 0,
\end{equation} 
where $D>0$ is a scale parameter, named diffusion constant, and $0<H<1$ is a parameter known as the Hurst exponent. The process was first introduced by Mandelbrot and Van Ness in \cite{mandelbrot}. It is a generalization of the Brownian motion, since for $D=1/2$ and $H=1/2$ it reduces to a standard Wiener process. We remark that fBm has the property of self-similarity, that is, for any choice of $a>0$,
\begin{equation*}
\{B_H(at), \; t\geq 0\} \stackrel{d}{=} \{a^H B_H(t), \; t \geq 0\},
\end{equation*}
meaning that the two processes are equal in distribution. The parameter $H$ is therefore also referred to as the scaling exponent or fractal index of the process. \par
The model (\ref{model}) is not stationary, since fBm is not a stationary process itself. The process of the increments of $B_H(t)$, defined by
$Z(t)=B_H(t+1)-B_H(t)$ for all $t \geq 0$,
is however stationary. This is known as fractionary Gaussian noise (fGn). \par
Introducing the process $Z(t)$ leads to another well-known property of fBm, that is the long-range dependency (LRD). As pointed out in \cite{dieker}, a stationary process $X(t)$ has long-range dependence (or is a long memory process) if its autocorrelation function, defined as
\begin{equation*}
\rho(k)=\frac{\textnormal{Cov}(X(t), X(t+k))}{\textnormal{Var}[X(1)]},
\end{equation*} 
satisfies the condition
\begin{equation*}
\sum_{k=-\infty}^{+\infty} \rho(k) = \infty.
\end{equation*}
This definition implies that the autocorrelation of the process decays slowly over time, making the sum of the autocorrelations divergent; therefore, if $X(t)$ is a LRD process, then
\begin{equation} \label{longmemory}
\lim_{k\rightarrow + \infty} \frac{\rho(k)}{c  k^{-\alpha}}=1,
\end{equation}
where $c>0$ and $0<\alpha<1$ are constants. This definition states that the decay of the autocorrelation function is power-like, hence slower than exponential. In the case of fBm, long-range dependence can be seen looking at the increments in $Z(t)$. Moreover, the parameter $\alpha$ is related to the Hurst exponent through the equation $\alpha=2H-2$, evidencing that the value of the Hurst exponent can be used to determine the nature of the process. We recall that (see also \cite{dieker} and \cite{li})
\begin{itemize}
\item if $\frac{1}{2}<H<1$, then the increments of the process are positively correlated, making the process persistent, i.e. likely to keep the trend exhibited in the previous observations;
\item if $0<H<\frac{1}{2}$, the increments of the process are negatively correlated and the process is counter-persistent, i.e. likely to break the trend followed in the past. 
\end{itemize} \par
A relation similar to (\ref{longmemory}), but concerning the frequency domain of the time series, is shown in Section 3, where it is used for the estimation of a parameter of the process.
\section{The deterministic component}
\label{sec:2}
\begin{figure}[!t]
\centering
\includegraphics[scale=0.6]{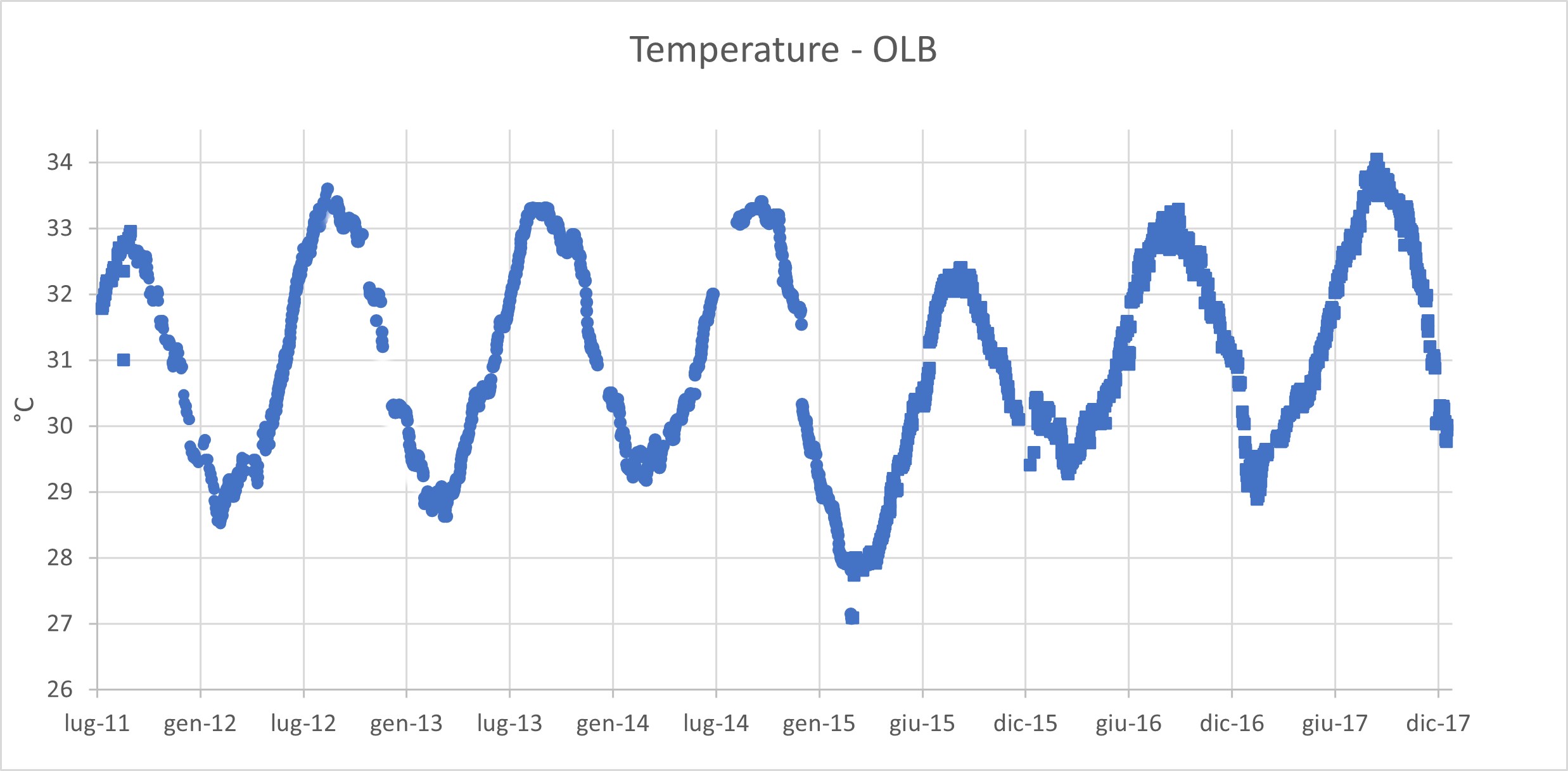}
\caption{Scatter plot of the data series $X(t)$ for the temperatures of the surface soil of the Campi Flegrei caldera in the Monte Olevano (NA) site. \label{fig:1}}
\end{figure}

The model (\ref{model}) is adopted to analyse the time series of the recorded temperatures of the Campi Flegrei caldera soil, cf. Sabbarese et al. \cite{0}. The initial data set consists of $\hat{N}+1=2005$ observations collected at the times indicated by $t_0, t_1, \ldots, t_{\hat{N}}$ and shown in the scatter plot in Figure \ref{fig:1}. 

\begin{figure}[!h]
    \centering
    \includegraphics[scale=0.35]{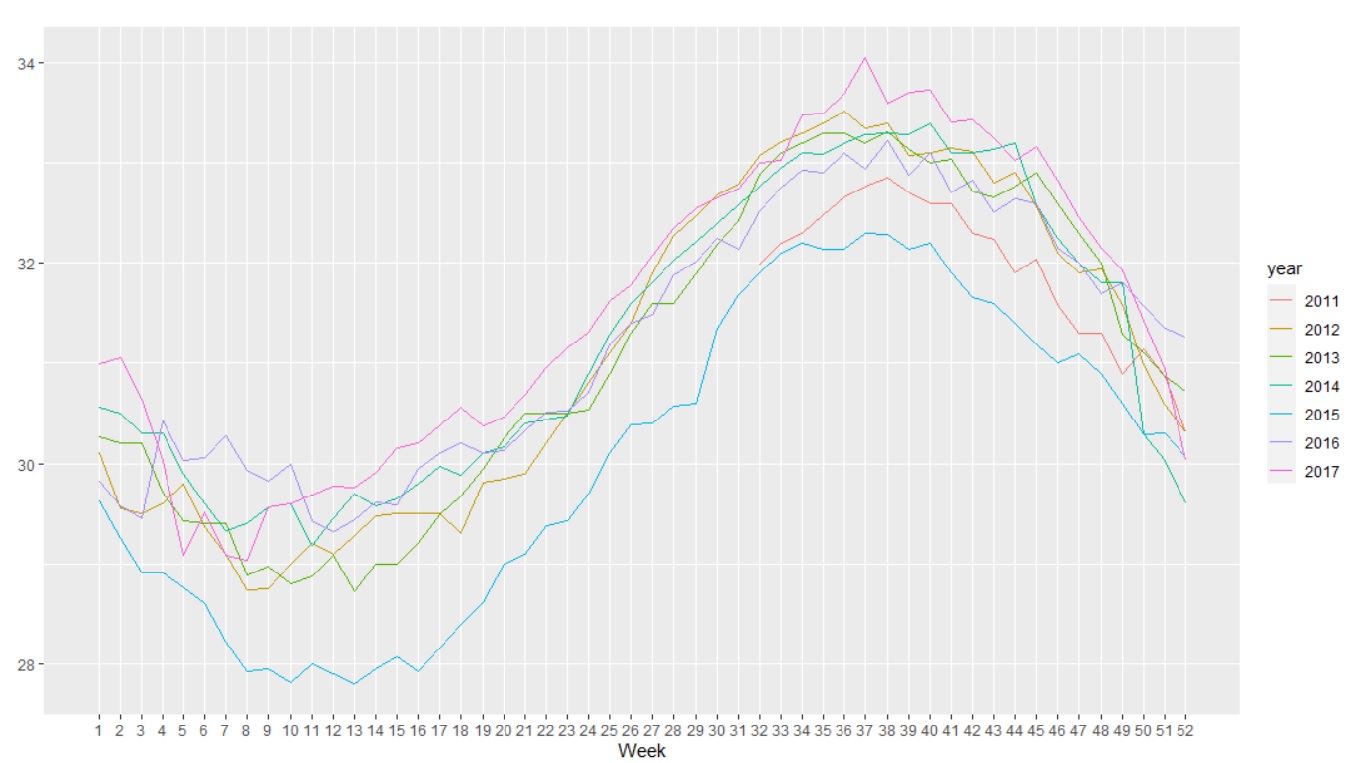}
    \caption{Seasonal plot of the temperatures described by $X(t)$. This is obtained by taking weekly spaced values from the time series $X(t)$ and representing each year of observation (2011-2017) separately.}
    \label{fig:seasonal}
\end{figure}

The observations are not equally spaced time-wise, since the data series was collected daily with lack of a quite large number of observations. As shown in the plot in Figure \ref{fig:seasonal}, a seasonal trend is clearly visible in the dataset, with the temperatures increasing during spring and summer in each year, and then starting to decrease around September. For this reason, the deterministic trend $r(t)$ is constructed by alternating segments with opposite slopes, obtained from the data through the OLS method.
The first and the last measurement of the dataset, which are also the extremes of the curve, are given by 
\begin{align*}
t_0=07/29/2011, \qquad X(t_0)=31.78; \\
t_{\hat{N}}=12/31/2017, \qquad X(t_{\hat{N}})=29.94.
\end{align*}
To obtain the points in which the curve changes its slope, an heuristic method based on linear regression is used, developed upon the methodologies illustrated by Hudson in \cite{hudson}. First, for the dataset, we detect the local minima $m_i, \;i=1,\ldots,6$ and maxima $M_i, \; i=1,\ldots,7$, as shown in Table \ref{tab:1}.

\begin{table}[H] 
\caption{Local maxima $M_i$ and minima $m_i$ for each subset of the dataset and the corresponding dates $\tau_i$ (mm/dd/year). \\ \label{tab:1}}
\centering
\begin{tabular}{|c|c|c|c|}
\hline
$\tau_i$	& $M_i=X(\tau_i)$ (\textdegree{}C)	& $\tau_i$ & $m_i=X(\tau_i)$ (\textdegree{}C) \\
\hline

 09/16/2011 & 31.78 & 02/14/2012 & 28.68  \\
 08/25/2012 & 33.6 & 03/22/2013 & 28.63 \\
 09/11/2013 & 33.31 & 03/04/2014 &29.17 \\
 09/22/2014 & 33.4 &  02/27/2015 & 27.08 \\
 09/05/2015 & 32.4 &  03/11/2016 &29.27 \\
 09/19/2016 & 33.29  & 02/03/2017 & 28.88 \\
 08/30/2017 & 34.04 & & \\

\hline
\end{tabular}
\end{table}

The time $t_0$ is used as the first extreme of the first segment of the piecewise linear curve. Subsequently, the slope $\alpha_j$ and the intercept of the regression line between $\tau_0$ and the first local maxima $M_1$ are estimated using the OLS method. The corresponding coefficient of determination $R^2$ is also calculated. The second step consists of repeating the operation, choosing the second extreme among the 5 values that precede and the 5 ones that follow $M_1$. Hence, among the 11 possible choices, we choose the regression line that maximises $R^2$. Table \ref{tab:2} shows the values obtained for each endpoint.

\begin{table}[H] 
\caption{Slopes of the segments obtained by linear regression between initial time $t_0$ and a neighbourhood of $M_1$, with their corresponding coefficients of determination. The slope is chosen such that $R^2$ is the highest, as underlined.  \\ \label{tab:2}}
\centering
\begin{tabular}{|c|c|c|}
\hline
$t_j$	& $\alpha_j$	& $R^2$\\
\hline
09/11/2011 & 0.0219  & 0.9302\\
09/12/2011 & 0.0217 & 0.9326\\
09/13/2011 & 0.0215 & 0.9341\\
09/14/2011 & 0.0213 & 0.9369\\
09/15/2011 & 0.0213 & 0.9405\\
$\underline{09/16/2011}$ & $\underline{0.0213}$ & $\underline{0.9442}$\\
09/23/2011 & 0.0199 & 0.9021\\
09/24/2011 & 0.0190 & 0.8787\\
09/25/2011 & 0.0182 & 0.8574\\
09/26/2011 & 0.0174 & 0.8369\\
09/27/2011 & 0.0162 & 0.7880\\
\hline
\end{tabular}
\end{table}

Then, the same method is applied for the next 12 segments, in which the first initial time is determined by the previous step. The last regression line is finally calculated using $t_{\hat{N}}$ as the endpoint. We denote by $c_i>0$ and $v_i<0$ the alternating slopes obtained through the regression model. The endpoints $\theta_i$ for each interval, as well as the estimated slopes and the coefficients of determination, are reported in Table \ref{tab:3}. \par
\begin{figure}[!h]
\centering
\includegraphics[scale=0.6]{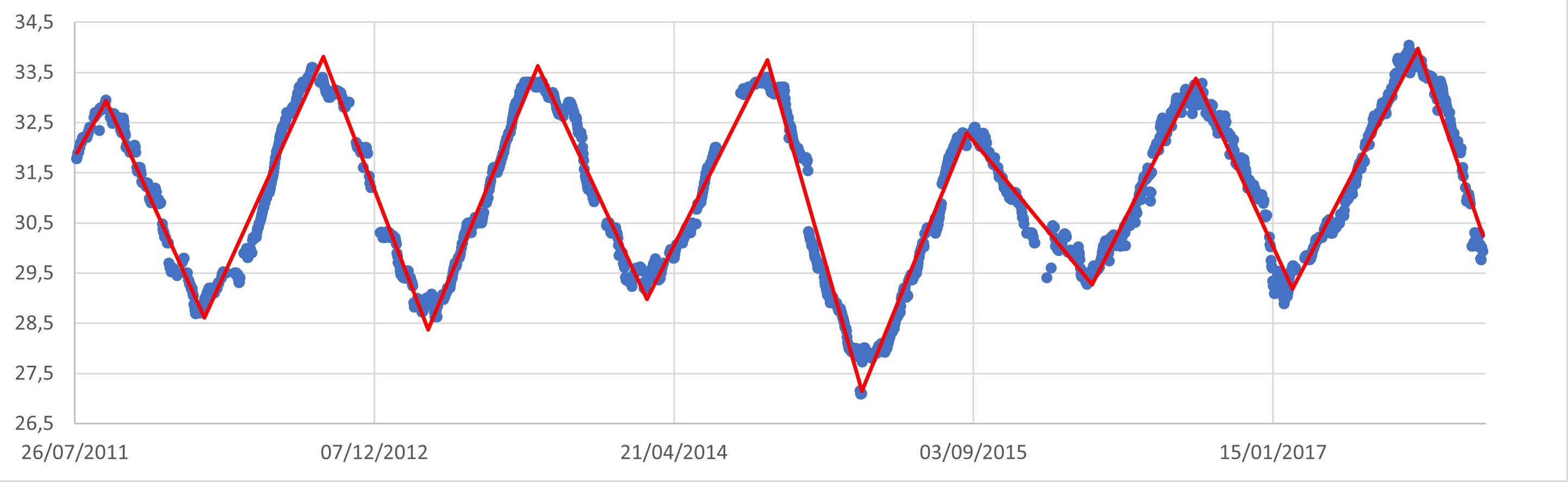} \\
\caption{The scatter plot of the data series $X(t)$ and the linear approximation of the deterministic trend $r(t)$. \label{fig:2}} 
\end{figure}

 
The values obtained by means of this procedure allow us to identify the deterministic component $r(t)$ of the model (\ref{model}). The values of $R^2$ confirm that the regression lines provide a good fitting of the data in each interval. The scatter plot and the resulting linear model are shown in Figure \ref{fig:2}.
\begin{table}[H] 
\caption{Estimated slopes $c_i$ and $v_i$, and corresponding coefficients of determination, evaluated for each subset $[\theta_{i-1},\theta_i]$ of the data series, with $\theta_0=t_0=07/29/2011$. \\ \label{tab:3}}

\centering
\begin{tabular}{|c|c|c @{\hspace{0.5cm}}| @{\hspace{0.5cm}}c|c|c|}
\hline
$\theta_i$	& $c_i$	& $R^2$ & $\theta_i$ & $v_i$ & $R^2$\\
\hline
09/16/2011 & 0.0213 & 0.9442 & 02/27/2012 & -0.0264 & 0.9709 \\
09/13/2012 & 0.0262 & 0.9356 & 03/07/2013 & -0.0311 & 0.9722\\
09/06/2013 & 0.0287 & 0.9818 & 03/07/2014 & -0.0256 & 0.9573 \\
09/24/2014 & 0.0237 & 0.9701 & 03/01/2015 & -0.0418 & 0.9635 \\
08/23/2015 & 0.0294 & 0.9659 & 03/19/2016 & -0.0145 & 0.9287\\
09/08/2016 & 0.0238 & 0.9619 & 02/16/2017 & -0.0261 & 0.9357\\
09/14/2017 & 0.0228 & 0.9674 & 12/31/2017 & -0.0346 & 0.9153\\
\hline
\end{tabular}
\end{table}

\section{Analysis of the stochastic component}
\label{sec:3}
\begin{figure}
\centering
\includegraphics[scale=0.7]{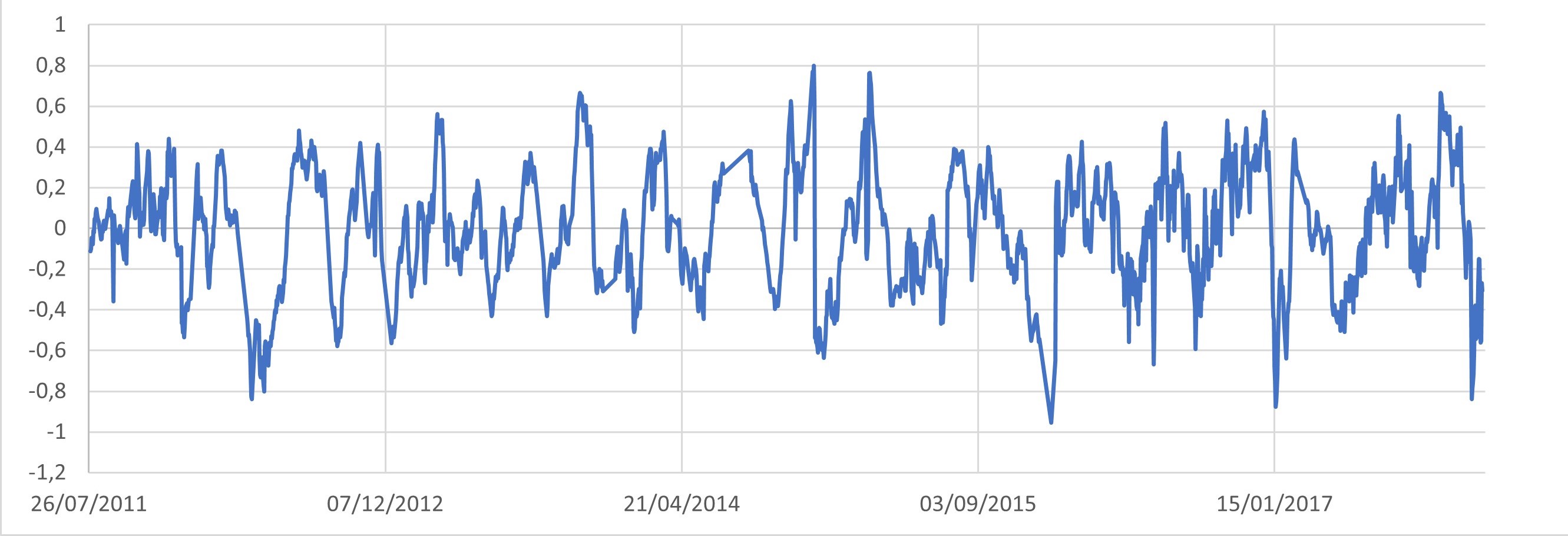}
\caption{Plot of the time series $B_H(t)=X(t)-r(t)$. \label{fig:3}}
\end{figure}

In this section we focus on the investigation concerning the nature of the time series $B_H(t)=X(t)-r(t)$. To this aim, some preparatory interpolations are performed on the dataset as customary. Since the times of the initial series are not equally spaced, we collect the missing data by performing a linear interpolation, covering the whole observation period. The prepared dataset obtained after this phase consists of $N+1=2348$ observations. A plot of $B_H(t)$ is shown in Figure \ref{fig:3}.
\par
\begin{figure}
    \centering
    \includegraphics[scale=0.5]{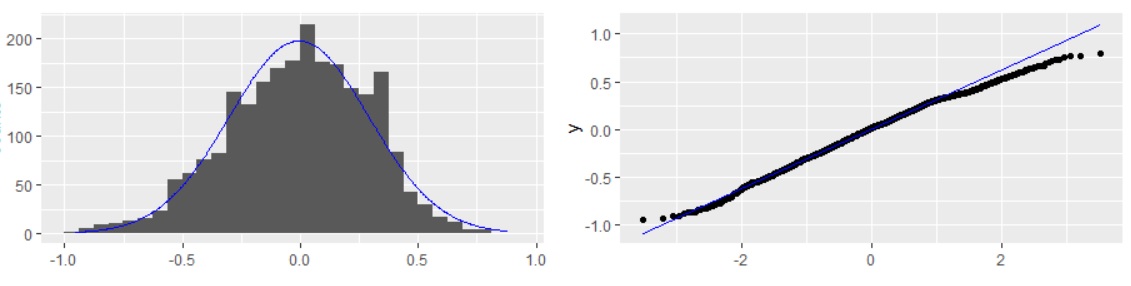}
    \caption{Histogram and qq-plot of the discrete time series $B_H(t)$.}
    \label{fig:histqq}
\end{figure}

 
As a preliminary study of the nature of the process $B_H(t)$, we represent the time series through an histogram and a qq-plot, as shown in Figure \ref{fig:histqq}. The obtained plots suggest to classify the stochastic component among Gaussian processes, thus justifying the following testing.
\subsection{Testing for Brownian motion}\label{sec:3.1}
 With reference to model (\ref{model}),  hereafter we perform a test on the process $B_H(t)$, 
 aiming to assess its suitable nature. With reference to the Brownian motion $B(t)$,
we consider the hypothesis  
\begin{align*}
& \mathcal{H}_0: B_H(t)=\sigma B(t), \\
& \mathcal{H}_1: B_H(t) \textnormal{ is a confined or directed diffusion}.
\end{align*}
By adopting the test proposed by Briane et al.\ in \cite{2}, 
the asymptotic region of acceptance for $\mathcal{H}_0$ is given by 
\begin{equation*}
\left\{ q \left( \frac{\alpha}{2} \right) \leq \frac{S_D^N}{\hat{\sigma}\sqrt{t_N}} \leq q \left(1-\frac{\alpha}{2} \right) \right\},
\end{equation*}
where
\begin{equation*}
S_D^N=\max_{j=1,\dotso,N} \vert B_H(t_j)-B_H(t_0) \vert,
\end{equation*}
\begin{equation*}
\hat{\sigma}=\left\{\frac{1}{N}\sum_{j=1}^N \frac{[B_H(t_j)-B_H(t_{j-1})]^2}{t_j-t_{j-1}}\right\}^{1/2},
\end{equation*}
and $q(\alpha)$ is the quantile of
\begin{equation*}
\sup_{0\leq s \leq 1} \vert B(s)-B(0) \vert.
\end{equation*}
The extremes of the acceptance interval for the significance level $\alpha=0.05$ are given by 
\begin{equation*}
q(0.025)=0.834, \qquad q(0.975)=2.940,
\end{equation*}
while the value obtained for the test statistic is $0.0787$. The null hypothesis of the process $B_H(t)$ being a Brownian motion is therefore rejected. This justifies the further analysis conducted hereafter. \par
\bigskip

Let us now conduct an investigation on $B_H(t)$ to verify whether it might follow a fBm or fGn trend. Consequently, we also estimate the Hurst exponent $H$ of the process.
To this aim, we apply the theoretical procedure consisting of evaluating the Fourier spectrum $S(f)$ of the data and studying its behaviour with respect to the frequencies $f$. In fact, a relation analogous to (\ref{longmemory}) for long memory processes can be presented in the frequency domain of the time series. In particular, for a fBm or a fGn process, the estimated Fourier spectrum $S(f)$ and the frequencies $f$ are asymptotically evaluated as
\begin{equation} \label{eq1}
S(f)=S(f_0)\, f^{-\beta},
\end{equation}
where both $S(f_0)$ and $\beta$ are constants. The coefficient $\beta$ is linked to the Hurst exponent $H$ by different equations, depending on the nature of the underlying process. Indeed, for a fGn (fBm) process, Eq.\,(\ref{eq1}) holds with $\beta=2H-1$ ($\beta=2H+1$), cf. \cite{roume}, \cite{taqqu} and \cite{flandrin}. Recalling that $ 0 < H < 1$, 
\begin{enumerate}
    \item[(i)]  if $-1<\beta<1$ then the time series can be identified as a realization of a fractional Gaussian noise, while 
    \item[(ii)] if $1<\beta<3$ then the data series represents a sample path of a fractional Brownian motion. 
\end{enumerate}
From (\ref{eq1}) we get
\begin{equation*}
\log S(f)=\log S(f_0) - \beta \log f,
\end{equation*}
so that an estimation of the coefficient $\beta$ con be obtained by plotting on a log-log scale the frequencies and the values of $S(f)$ and then taking the opposite value of the slope of the least squares line as estimation. 
\par

\begin{figure}[t]
\centering
\includegraphics[scale=0.7] {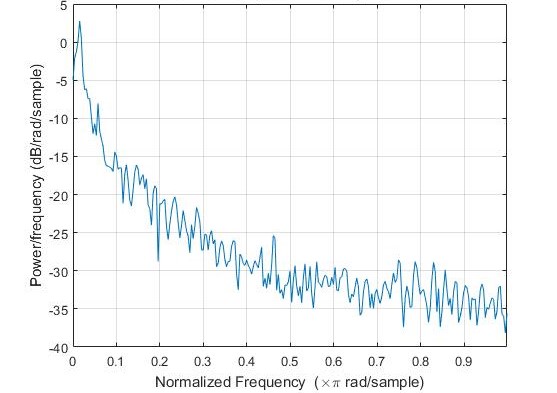}
\caption{Modified Welch periodogram for the data series $B_H(t)$, plotted for different values of the frequency. \label{fig:4}}
\end{figure}

To attain an estimation of function $S(f)$, we evaluate the periodogram of the data series, which for $B_H(t)$ is defined by
\begin{equation*}
    I(f)=\left\lvert \, \frac{1}{2\pi N}\sum_{i=0}^N B_H(t_i)e^{i f t_i}\, \right\lvert ^2.
\end{equation*}
In order to achieve the best possible approximation for the Fourier spectrum, many modified versions of the periodogram have been devised. In particular, we choose the estimation of $S(f)$ using the modified periodogram method suggested by Welch in \cite{4}. The technique consists in dividing the signal into overlapping segments and averaging the modified periodograms calculated in each window to obtain the final result. The Welch's modified periodogram is shown in Figure \ref{fig:4}. \par

\begin{figure}[t] \
\centering
\includegraphics[scale=0.5] {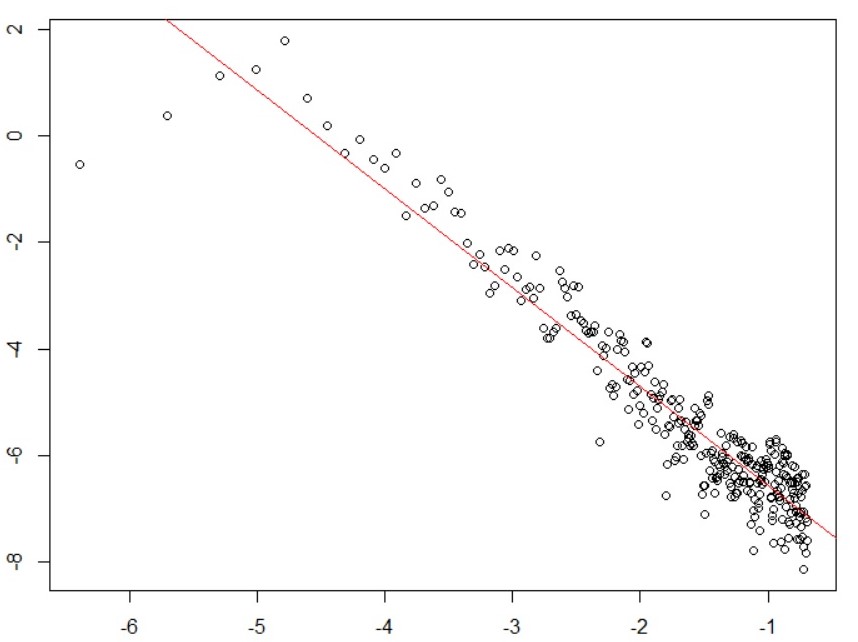}
\caption{Log-log plot for the Welch's periodogram of the data series with respect to frequency. The slope of the least squares line estimates the coefficient $\beta$. \label{fig:5}}
\end{figure}

 Hence, by applying a log-log transformation on the frequencies and the estimated periodogram, we obtain the plot shown in Figure \ref{fig:5}. The estimate of $\beta$ obtained as the slope of the regression line is given by
 \begin{equation*}
     \beta=1.853,
 \end{equation*}
 with corresponding $R^2=0.907$. Since the coefficient $\beta$ belongs to the interval $(1,3)$, we can conclude that the data series follows a fBm trend. This also leads to an estimation of the Hurst parameter as follows:
 \begin{equation}
     \hat{H}=\frac{\beta-1}{2}=0.427.
     \label{hurst}
 \end{equation}

 In order to complete the estimation of the parameters for $B_H(t)$, we also evaluate the diffusion constant $D$ of the process. This is done directly from the data, calculating the sample variance of the increments $B_H(t)-B_H(t-1)$ of the data series. The estimate obtained so far is
\begin{equation}
\hat{D}=0.0846.
\label{scale}
\end{equation}

To further confirm the hypothesis suggested from the periodogram, we perform an analysis of the residuals of the time series making use of simulations of fBm. To this aim, we denote with $B_{\hat{H},\hat{D}}(t)$ a simulated sample path of a fractional Brownian motion process with Hurst exponent and scale parameter estimated as in (\ref{hurst}) and (\ref{scale}). Then, the residual 
\begin{equation*}
    z(t)=X(t)-r(t)-B_{\hat{H},\hat{D}}(t)
\end{equation*} 
is evaluated by means of the residuals plot. Moreover, the normality of the residuals is assessed hereafter through two tests: the Shapiro-Wilk and the Robust Jarque-Bera test for Gaussianity. 
\par
\subsection{Shapiro-Wilk test}
\begin{figure} [t]
    \centering
    \includegraphics[scale=0.65]{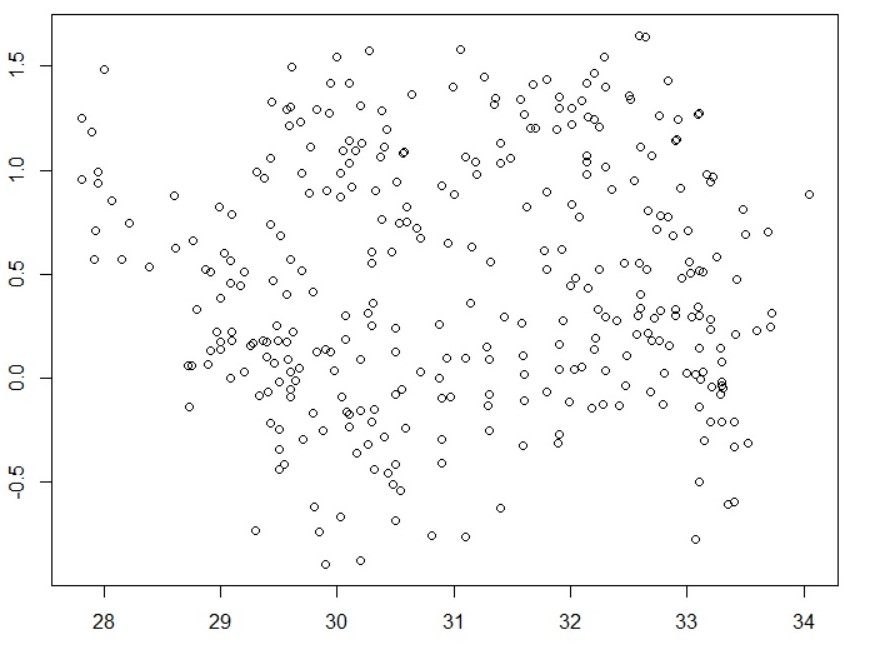}
    \caption{Plot of the residuals series $\tilde{z}(t)$ with respect to the temperatures $\tilde{X}(t)$, equally spaced with one observation for each 7 days.  \label{fig:6}}
\end{figure}

Since the test statistic $W$ of the Shapiro-Wilk test tends to detect even small departures from the null hypothesis when the sample size is sufficiently large, it is convenient reducing the sample to a smaller size. 
Hence, from now on we consider a subset $\tilde{X}(t)$ of the main dataset, which contains weekly spaced observations. After simulating a sample path $\tilde{B}_{\hat{H},\hat{D}}$ of a fBm of length $M=\lfloor (N+1)/7 \rfloor $, we plot the residual series 
\begin{equation*}
    \tilde{z}(t)=\tilde{X}(t)-\tilde{r}(t)-\tilde{B}_{\hat{H},\hat{D}}(t)
\end{equation*} 
with respect to the values of $\tilde{X}(t)$, see  Figure \ref{fig:6}. Notice that the residuals do not seem to follow a specific path, but appear to be distributed randomly with respect to $\tilde{X}(t)$.

\begin{figure}[!h]
\begin{minipage}[t]{0.45\linewidth}
            \centering
             \includegraphics[scale=0.3]{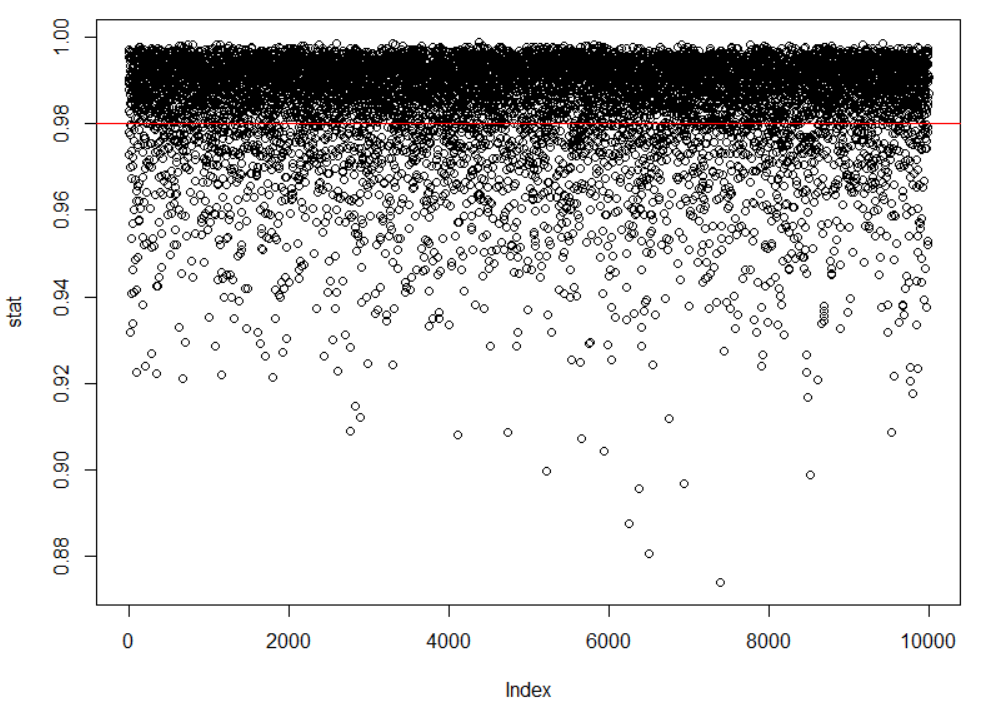}
             \vspace{-0.75cm}
    \caption{The values of the statistic $W_i$ evaluated according to Eq.\ (\ref{w}). 
    \label{fig:7}}
            \label{fig:a}
        \end{minipage}
        \hspace{1.2cm}
        \begin{minipage}[t]{0.45\linewidth}
        \vspace{-4.45cm}
            \centering
            \includegraphics[ scale=0.275]{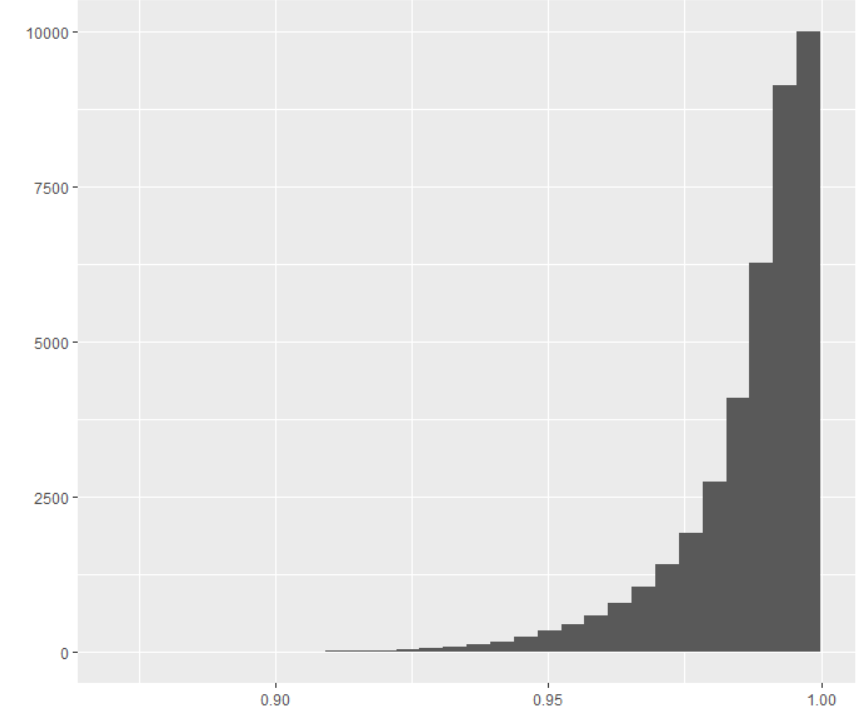}
            \caption{\footnotesize Cumulative histogram for the $W_i$ statistic. \\
            }
            \label{fig:7b}
        \end{minipage}
    \end{figure}

The fBm path $\tilde{B}_{\hat{H},\hat{D}}(t)$ is then simulated again for a total of $n=10^4$ iterations. For each  simulation we evaluate the  residuals series $\tilde{z}_i(t)$ and first perform a Shapiro-Wilk test, for $i=1,\ldots,n$. 
Hence, we consider the collection of test statistics
\begin{equation}
    W_i=\frac{\sum_{j=1}^M a_j \tilde{z}_{i}(t_{(j)})}{\sum_{j=1}^M (\tilde{z}_i(t_j)-\mu_{i})^2}, \qquad i=1,\ldots, n,
    \label{w}
\end{equation}
where $W_i$ refers to the $i$-th simulation, and where 
$\mu_{i}$ represents the sample mean of $\tilde{z}_i(t)$. The coefficients $a_j$ --as well known--  are related to some moments of the order statistics of i.i.d. random variables sampled from the standard normal distribution.
The values of $W_i$ as in Eq.\ (\ref{w}) range between 0 and 1. 
The $i$-th test rejects the Gaussian hypothesis when $W_i$ is close to 0, hence we accept the null hypothesis if $W_i \geq 0.98$, for $i=1,\ldots,n$. 
As shown in Figure \ref{fig:7}, large proportion of the values of  $W_i$ is greater than $0.98$. This result has been observed for more than 70\% of the values of $W_i$ even for various repetitions of the $n$-simulation procedure (cf. \cite{royston1}, \cite{royston2}). This is also confirmed by the cumulative histogram of $W_i$ shown in Figure \ref{fig:7b}. The results of the test suggest the acceptance of the Gaussianity hypothesis. \par

\subsection{Robust Jarque-Bera test}
Let us now perform a modified version of the Jarque-Bera test. The statistic for the original Jarque-Bera test for a sample of size $n$ is given by (cf. \cite{jbtest})
\begin{equation*}
    JB=\frac{n}{6}\left( \frac{\hat{\mu}_3}{\hat{\mu}_2^{3/2}}\right)^2+\frac{n}{24}\left(\frac{\hat{\mu}_4}{\hat{\mu}_2}-3\right)^2,
\end{equation*}
which is a combination of the sample skewness $\left(\frac{\hat{\mu}_3}{\hat{\mu}_2^{3/2}}\right)$ and the sample kurtosis $\left(\frac{\hat{\mu}_4}{\hat{\mu}_2}-3\right)$ obtained from the data through the estimates $\hat{\mu}_i$ of the central moments, $i=2,3,4$. If the data series is distributed normally, one has that $JB\sim \chi^2(2)$ asymptotically. Thus, under the null hypothesis of the time series being Gaussian, both the expected skewness and the expected kurtosis are 0. Therefore, higher values of the $JB$ statistic lead to the rejection of the Gaussianity hypothesis. \par
For our testing procedure, we follow  the modified version of the test proposed by Gel and Gastwirth in \cite{gel}, which is known as the Robust Jarque-Bera (RJB) test. The modification consists in substituting the estimate of the spread, formerly $\hat{\mu}_2$, with the average absolute deviation from the sample median, henceforth denoted as $\textnormal{Med}$, given by
\begin{equation*}
    J_M=\frac{\sqrt{\pi/2}}{M} \sum_{j=1}^M |\tilde{z(t_j)}-\textnormal{Med}|,
\end{equation*}
where $M$ refers to the sample size. This substitution makes the statistics less influenced by outliers. Therefore, the statistic becomes
\begin{equation*}
    JB_M=\frac{M}{C_1}\left( \frac{\hat{\mu}_3}{J_M^{3}}\right)^2+\frac{M}{C_2}\left(\frac{\hat{\mu}_4}{J_M^4}-3\right)^2,
\end{equation*}
where $C_1$ and $C_2$ are positive constants estimating moments of $J_M$, obtained through Monte Carlo simulations. Moreover, the $p$-values for the RJB test are obtained by use of $k=10000$ Monte Carlo simulations instead of the $\chi^2(2)$ distribution, thus obtaining a better approximation (cf.\ \cite{deb}).  
\par
To perform the RJB test, we repeat 10 blocks of $n$ simulations with three different numbers of iterations ($n=100,500$ and 1000). For each simulation, the residual series \begin{equation}
    \{\tilde{z}_i(t_j), \; j=1,\ldots, M\}_{i=1,\ldots,n}
    \label{ztilde}
\end{equation} is tested. The results, reported in Table \ref{tab:4}, show compliance with the Shapiro-Wilk test and  lead to the acceptance of the Gaussianity hypothesis for the residuals series. 
\begin{table}[!h] 
\caption{Percentages of acceptable values for the residuals series (\ref{ztilde}), tested with the RJB method, repeated for $h=1,\ldots,10$, with significance level $\alpha=0.02$ and different values of $n$. \\ \label{tab:4}}
\centering
\begin{tabular}{|c|c|c|c|}
\hline
$h$ & $n=100$	& $n=500$ & $n=1000$\\
\hline
1 & 64\% & 66.8\% & 68.0\%  \\
2 & 76\% & 72.2\% & 69.9\% \\
3 & 68\% & 70.0\% & 66.1\% \\
4 & 66\% & 69.4\% & 71.3\% \\
5 & 70\% & 71.0\% & 70.5\%\\
6 & 68\% & 71.4\% & 71.1\%\\
7 & 68\% & 70.8\% & 70.5\%\\
8 & 71\% & 67.0\% & 69.9\%\\
9 & 70\% & 73.0\% & 68.1\%\\
10 & 70\% & 69.4\% & 71.5\%\\
\hline
\end{tabular}
\end{table}
%

\section{Statistical test for the model}

To further confirm the results obtained in Section \ref{sec:3}, we perform a statistical test on the whole $N$-term data series for the hypothesis of $B_H(t)$ being a fractional Brownian motion with parameters $\hat{H}=0.427$  and $\hat{D}=0.0846$. To this aim, we use the inference test presented in Sikora \cite{6}, based on the detrending moving average (DMA) of the data. The hypotheses for the test are
\begin{center}
$\mathcal{H}_0: \{B_H(t_1),B_H(t_2), \ldots, B_H(t_N)\}$ \textit{ is a trajectory of fBm with parameters $\hat{H}$ and $\hat{D}$},
\end{center}
\begin{center}
$\mathcal{H}_1: \{B_H(t_1),B_H(t_2), \ldots, B_H(t_N)\}$ \textit{ is not a trajectory of fBm with parameters $\hat{H}$ and $\hat{D}$}.
\end{center}
The corresponding test statistic, for fixed $m>1$, is given by
\begin{equation}\label{sigma2}
S^2(m)=\frac{1}{N-m} \sum_{j=m}^N (B_H(t_j)-\overline{B}_H^m(t_j))^2,
\end{equation}
where 
$\overline{B}_H^m(t_j)$ denotes the moving average of the $m$ observations $B_H(t_{j-i}), \; i=0,1,\ldots, m-1$. 

The $p$-value for this test is defined by an infinite series, so it needs to be computed as an empirical quantile, obtained from series of generalized chi-squared random variables. Recalling Step 5 in Section 3 of Sikora \cite{6}, the $p$-value of the test is calculated as
\begin{equation} \label{pvalue}
p=\frac{2}{L}\,\min \{\#(\sigma^2_l(m)<S^2(m)), \#(\sigma^2_l(m)>S^2(m))\}.
\end{equation}
In this formula, $L$ is the number of samples generated for the chi-squared distribution. For a good estimate of $p$ we set $L=1000$. Moreover, in Eq.\ (\ref{pvalue}), we consider
\begin{equation*}
\sigma^2_{l}(m)=\frac{1}{N-m} \sum_{j=1}^{N-m+1} \lambda_{j}(m) U_j^l, \qquad l=1,2,\ldots, L,
\end{equation*}
where $\textnormal{\textbf{U}}^l=\{U^l_1,U^l_2, \ldots, U^l_{N-m+1}\}$ is the $l$-th chi-square sample and $\lambda_j(m)$ is the $j$-th eigenvalue of the sample matrix 
\begin{equation*}
\tilde{\Sigma}=\{\mathbb{E}[B_H(t_j)B_H(t_k)]; \; j,k=1,2,\ldots, N-m+1\}. 
\end{equation*}

\begin{figure}[t]
\includegraphics[scale=0.25]{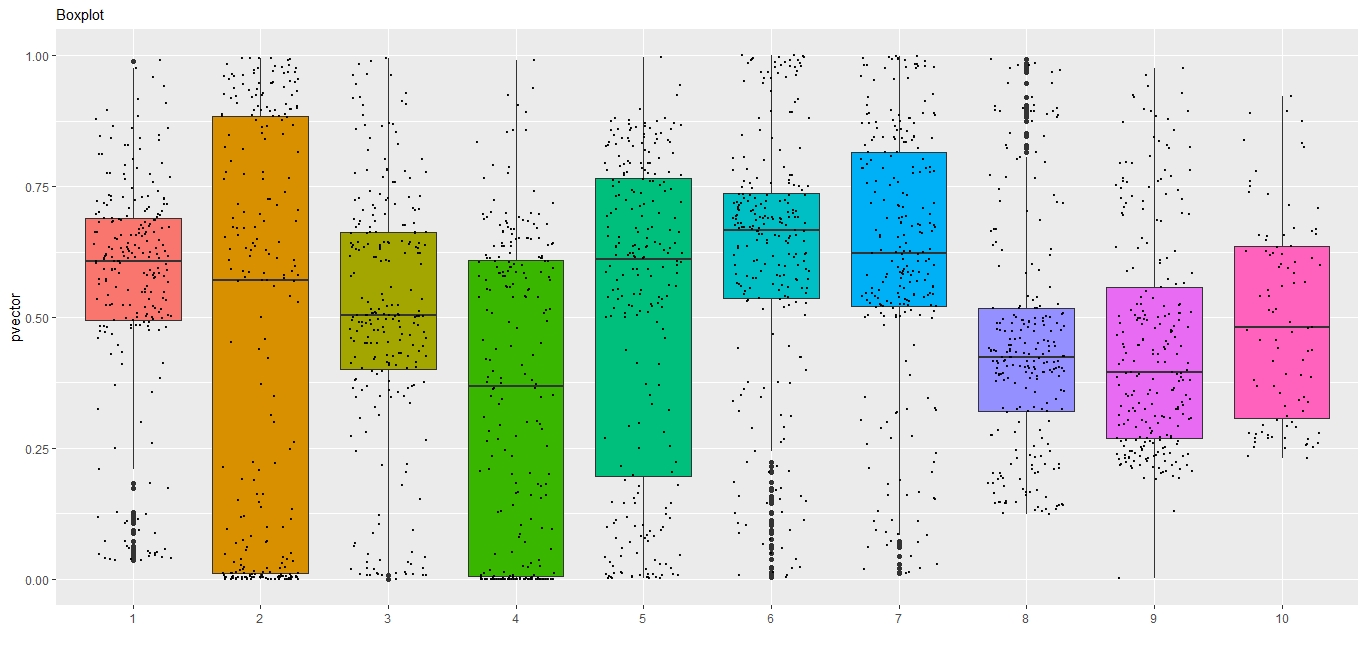}
\caption{Box plot for the values of $p$ obtained in the 10 subsets of the data. \label{fig:8}}
\end{figure}
\par
Recalling (\ref{sigma2}), since the search for an optimal value for $m$ is still an open question, we divide the data series into 10 subsets of about 250 observations each, in order to adapt the procedure even to cases with large number of observed data. Then we run the test in every subset for each suitable value of $m$. 
\par 
After that, we collect the $p$-values and create a box plot with jitter for each subset, in order to determine the outlier values for $p$, cf.\ Figure \ref{fig:8}.  
\par
The significance level considered for this test is $\alpha=0.02$. Therefore the results in each subset are collected and divided into three categories: \\
\begin{enumerate}
    \item[(i)] for $p\leq 0.02$, the null hypothesis is rejected,\\
    \item[(ii)] for $0.02<p\leq 0.05$ the values are considered as "warning", \\
    \item[(iii)] for $p> 0.05$ the hypothesis $\mathcal{H}_0$ cannot be rejected. 
\end{enumerate}
Table \ref{tab:5} shows the results, with the percentages of the number of rejected, warning and acceptable values over the total of $p$-values calculated in the 10 datasets. It can be observed that the null hypothesis cannot be rejected in almost 90\% of the cases covered, considering all the subsets  and all the different choices of $m$ for the test statistic. This allows to state the validity of the hypothesis that the time series $B_H(t)$ follows a fBm trend.

\begin{table}[!h] 
\caption{Number of cases in which the DMA statistic test has different outcomes for the data series $B(t)$. \\ \label{tab:5} }
\centering
\begin{tabular}{|c|c|c|c|c|c|c|}
\hline
 Set & Reject values & Perc. & Warning values & Perc. &  Acceptable values & Perc. \\ & ($p\leq 0.02$) &  & ($0.02 \leq p\leq 0.05$) & &  ($p>0.05$) \\
\hline
1 & 0 & 0\% & 12 & 4.86\% & 235 & 95.14\% \\ 
 2 & 70 & 28.23\% & 12 & 4.84\% & 154 & 62.10\% \\ 
 3 & 17 & 6.85\% & 6 & 2.42\% & 219 & 88.31\% \\
 4 & 70 & 28.23\% & 6 & 2.42\% & 166 & 66.94\% \\
 5 & 23 & 9.27\% & 10 & 4.03\% & 205 & 82.66\% \\
 6 & 4 & 1.61\% & 4 & 1.61\% &  236 & 95.16\% \\
 7 & 3 & 1.21\% & 3 & 1.21\% & 239 & 96.37\% \\
 8 & 0 & 0\% & 0 & 0\% & 248 & 100\% \\
 9 & 1 & 0.4\% & 0 & 0\% & 247 & 99.6\% \\
 10 & 0& 0\% & 0 & 0\% & 95 & 100\% \\
\hline
\end{tabular}
\end{table}

\section{Conclusions}
The performed analysis shows that the proposed stochastic model is suitable to fit the observed temperatures. Indeed,
the obtained results suggest that the coefficients obtained for the process $X(t)$  display high values for $R^2$, this ensuring the goodness of the fit. 
In conclusion, the study reveals that the evolution of the temperatures in the Campi Flegrei caldera can be modeled by the sum of a deterministic component, that represents the seasonal trend, and a fractional Brownian motion.
\par
Possible  future developments of the performed analysis can be oriented (i) to the prevision of the trend of the seasonal terms,  and (ii) to the study of other quantities of interest, such as the leaked radon. 

\paragraph{Acknowledgments}
The authors are members of the research group GNCS of INdAM (Istituto Nazionale di Alta Matematica). This work is supported in part by the Italian MIUR-PRIN 2017, project ‘Stochastic Models for Complex Systems’, No. 2017JFFHSH.

\end{document}